\documentclass[12pt]{article}
\usepackage{amsfonts,amssymb}

\def\N{\mathbb{N}}

\def\bq{ \begin{equation} }
\def\eq{ \end{equation} }
\def\ben{ \begin{eqnarray} }
\def\en{ \end{eqnarray} }

\def\frac#1#2{{#1\over #2}}

\def\on#1#2{\mathop{\vbox{\ialign{##\crcr\noalign{\kern2pt}
$\scriptstyle{#2}$\crcr\noalign{\kern2pt\nointerlineskip}
\kern-2pt$\hfil\displaystyle{#1}\hfil$\crcr}}}\limits}

\title{On integrability of the Kontsevich \\ non-abelian ODE system}
\author{Thomas Wolf\\ 
Brock University      
(Ontario, Canada) \\  
twolf@brocku.ca 
\\ tel (1) 905 688 5550 x 3803, fax (1) 905 378 5713
\vspace*{10pt} \\
Olga Efimovskaya \\
Moscow State University (Russia),\\
olga.efimovskaya@gmail.com}
\begin{document}

%%%%%%%%%%%%%%%%%%%%%%%%%%%%%%%%%%%%%%

\maketitle
\begin{abstract}
  We consider systems of ODEs with the right hand side being Laurent
  polynomials in several non-commutative unknowns. In particular,
  these unknowns could be matrices of arbitrary size. An important
  example of such a system was proposed by M. Kontsevich. We prove the
  integrability of the Kontsevich system by finding a Lax pair,
  corresponding first integrals and commuting flows. We also provide
  a pre-Hamiltonian operator which maps gradients of integrals for 
  the Kontsevich system to symmetries. \vspace{10mm}  \vspace{5mm} 
\end{abstract}

\newpage 

 %---------------------
\section{Introduction}

In connection with the theory of non-commutative elliptic functions,
M.\ Kontsevich \cite{MKpriv} considered the following discrete map
\begin{equation}\label{descr}
  u \rightarrow  u v u^{-1}, \qquad  v \rightarrow  u^{-1}+v^{-1} u^{-1},
\end{equation}
where $u,v$ are non-commutative variables (in particular, $n \times
n$-matrices of arbitrary size).  His numerical computer experiments
have shown that this map could be integrable (see \cite{vesel}). In
the abelian case the element
\begin{equation}\label{hh} h=u+v+u^{-1}+v^{-1}+u^{-1} v^{-1}\end{equation} 
is an integral for the mapping (\ref{descr}). The equation $h=const$
defines a family of elliptic curves.  In the non-abelian case the
element $h$ is transformed as $h \rightarrow u h u^{-1}$. It follows from
this formula that $trace(h^k)$ is a first integral of (\ref{descr})
for any natural $k$.

Kontsevich also observed that (\ref{descr}) is a discrete
symmetry of the following non-abelian ODE system:
\begin{equation}\label{inhom}
u_t=u v -u v^{-1} - v^{-1}, \qquad v_t=- v u + v u^{-1} + u^{-1}
\end{equation}
and conjectured that (\ref{inhom}) is integrable itself. 

Our paper is devoted to the system (\ref{inhom}). It belongs to the
class of systems of the form
\begin{equation}
\label{eq}
u_{t}=P_{1}(u,\,v), \qquad v_{t}=P_{2}(u,v),
\end{equation}
where $P_{i}$ are elements of the associative algebra ${\mathcal M}$ of all
non-commutative polynomials in $u,v,$ $u^{-1}, v^{-1}$ with constant scalar coefficients. 
The elements of ${\mathcal M}$ are called  {\it non-abelian Laurent polynomials}. 

In papers \cite{MikSok,Ef} integrable systems of type (\ref{eq}) with
$P_i$ being non-abelian polynomials in $u,v,$ were considered. The
existence of an infinite series of infinitesimal symmetries was taken
as a criterion for integrability. A similar approach to integrability
of evolutionary polynomial non-abelian PDEs was developed in
\cite{olsok}.  As far as we know integrable systems with non-abelian
Laurent right hand sides were not considered before.

In this paper we find a Lax representation with a spectral parameter
for system (\ref{inhom}). The corresponding Lax $L$-operator generates
infinitely many integrals of motion for (\ref{inhom}). They are
integrals for the discrete map (\ref{descr}) as well.  We also find a
pre-Hamiltonian operator that maps gradients of first integrals to
symmetries.  This proves that (\ref{inhom}) is integrable in the sense
of \cite{MikSok}.

\medskip 
\medskip 

\section{Symmetries and integrals} \label{SymInt}

Let us briefly recall the definitions from
\cite{MikSok} generalized to the Laurent case.

Let $x_{1},\dots, x_{N}$ be non-commutative variables. We consider ODE
systems of the form
\begin{equation}\label{geneq}
\frac{d x_{\alpha}}{d t}=F_{\alpha}(x_1,...,x_N,x_1^{-1},...,x_N^{-1}), 
\end{equation}
where $F_{\alpha}$ are Laurent polynomials. We denote by ${\mathcal M}$ 
the associative algebra of all Laurent polynomials. Formulas
(\ref{geneq}) together with 
$\frac{d(x_{\alpha}^{-1})}{d t} = - x_\alpha^{-1} 
 \frac{d x_{\alpha}}{d t} x_\alpha^{-1}$ 
define the corresponding derivation $D_t$ on ${\mathcal M}$.

An (infinitesimal) symmetry for (\ref{geneq}) is a system 
\begin{equation}\label{gensym}
\frac{d x_{\alpha}}{d \tau}=G_{\alpha}(x_1,...,x_N,x_1^{-1},...,x_N^{-1}), 
      \qquad G_{\alpha}\in {\mathcal M}
\end{equation}
compatible with (\ref{geneq}). Compatibility means that the derivations $D_t$
and $D_{\tau}$ corresponding to (\ref{geneq}) and (\ref{gensym}) commute.

The existence of an infinite series of symmetries is typical for integrable
non-abelian ODEs.  The simplest non-trivial symmetry for (\ref{inhom}) is
given by
\begin{eqnarray*}
u_{\tau}&=& - u v u - u v^{2} + u v + (v u)^{-1} + v^{-2} + u^{2}v^{-1} 
          - u v u^{-1} + u v^{-2}
+ (v u v)^{-1}+ u ( v u v)^{-1}, \\
v_{\tau}&=&\ \ \, v u v  +   v u^{2} - v u - (u v)^{-1} - u^{-2} - v^{2} u^{-1}   
          + v u v^{-1} - v u^{-2}
- ( u v u)^{-1}- v (u v u)^{-1}.
\end{eqnarray*}
The following conjecture is concerned with the dimensions of vector spaces
$S_k$ of all symmetries with right hand sides being polynomials in
$u,v,u^{-1},v^{-1}$ of degree $\le k.$

{\bf Conjecture:\vspace{3pt}}\\
${\rm dim}\, S_{4n}=2 n^2,\quad{\rm dim}\, S_{4n+1}=2 n^2+2 n,\quad{\rm dim}\, 
 S_{4n+2}=2 n^2+2 n+1,\quad{\rm dim}\, S_{4n+3}=2 n^2+4n+1$.

We have verified the conjecture for $S_k,$ $k=0,...,16.$ Notice that (\ref{inhom}) 
is invariant with respect to involution 
\begin{equation}\label{invol}
 \quad u\to v,\qquad v\to u
 \end{equation}
 and $t\to -t$. All known symmetries are either involution invariant up
 to $\tau \to -\tau$ or involution symmetric to another symmetry.  
In the remainder of the paper we will denote
 the involution $F|_{v \leftrightarrow u}$ of any Laurent polynomial
 $F$ by $\overline{F}$.

An element $I$ of ${\mathcal M}$ is called a {\it full integral} for
(\ref{geneq}) if $D_t(I)= 0$. In the matrix case this means that
every component of $I$ is an integral of motion. For equation (\ref{inhom})
the element 
\begin{equation}\label{II}
I=uvu^{-1} v^{-1}\end{equation}
is a full integral. This element is
an integral for the mapping (\ref{descr}) as well. We were unsuccessful in
finding more Laurent full integrals for (\ref{inhom})
(other than polynomials of $I, I^{-1}$). 

In the $q$-case, which is a specialization of our general non-abelian
situation, an additional full integral exists. Namely, consider the
associative algebra ${\mathcal Q}$ generated by $u,v$ with the
relation $u\,v=q \, v \,u,$ where $q$ is a fixed constant. In this
case the element
$$
J=u +q v +q u^{-1}+v^{-1}+u^{-1} v^{-1} 
$$
is a full integral for (\ref{inhom}). This integral is a $q$-deformation 
of the element (\ref{hh}).

Apart from (\ref{II}), equation (\ref{inhom}) has so called trace first
integrals. If $u$ and $v$ are matrices, these integrals are given by
traces of some Laurent polynomials. For instance, the traces of $h^k$,
where $h$ is given by (\ref{hh}) and $k\in \N$, are such integrals.
 
For (\ref{geneq}) with $x_\alpha$ being non-commutative symbols we 
define the $trace(a), \,\, a\in {\mathcal M}$ as an equivalence class. 
Two elements $a$ and $b$ of ${\mathcal M}$ are called equivalent iff
$a$ can be obtained from $b$ by cyclic permutations of factors in its
monomials.  In other words, the traces for elements of
${\cal M}$ are defined as the corresponding elements of the quotient space
${\cal M}/K,$ where $K$ is the vector space spanned by all commutators
in ${\cal M}$.  
If $a-b\in K$, we write $a \sim b$.

An element $\rho$ of ${\mathcal M}$ is called a {\it trace integral}
of (\ref{geneq}), if $D_t(\rho) \sim 0$.  Trace integrals $\rho_1$ and
$\rho_2$ are called equivalent if $\rho_1 - \rho_2 \sim 0$. By
definition, the degree of a trace integral is the minimal degree of
elements from the corresponding equivalent class.

Low degree trace integrals of (\ref{inhom}) can be found straightforwardly.
In particular, there are no integrals of degrees 1 and 3, $h$ is the only
integral of degree 2, three linearly independent trace integrals of degree 4 are
given by $h^2,I$ and $I^{-1}$. Detailed information of higher degree integrals
can be found in Section \ref{LaxPair}.

To generate infinite sequences of trace integrals for (\ref{inhom}) the 
following procedure can be applied. 

Suppose two Laurent polynomials  $H,A \in {\cal M}$ satisfy 
\begin{equation}\label{LAX}
D_t(H)=[A,\,H]
\end{equation}
where $D_t$ is the derivation corresponding to (\ref{inhom}). The
involution (\ref{invol}) generates the pair $\overline{A}, \overline{H}$ 
that is also satisfying (\ref{LAX}). It follows from (\ref{LAX}) that $H$ is a
trace integral for (\ref{inhom}).  Since $H^k$ also satisfy
(\ref{LAX}) for any natural $k$ the elements $H^{k}$ are trace
integrals as well. In the matrix case this is equivalent to the fact
that the spectrum of $H$ is preserved under the flow (\ref{inhom}).
Thus for any pair $(H_i,A_i)$ satisfying (\ref{LAX}) the elements
\begin{equation} \label{Hjk}
          \rho_{ik} =           H_i^{\, k}, \qquad {\rm and} \qquad  
\overline{\rho_{ik}}= \overline{H_i}^{\, k}
\end{equation}
are trace integrals for any natural $k$. 

It is easy to see that if $H,A$ satisfy (\ref{LAX}), then for any
invertible element $g\in {\cal M}$ the conjugation
\begin{equation} \label{gauge}
H\to g H g^{-1}, \qquad A\to g A g^{-1}+g_t g^{-1}
\end{equation}
leads to another pair satisfying (\ref{LAX}). It is clear that
conjugated pairs produce the same trace integrals (\ref{Hjk}).

Apart from $h$ given by (\ref{hh}) satisfying the relation (\ref{LAX})
with $A= -u^{-1}-v$ we found several more low degree pairs $(H_i,A_i)$
satisfying (\ref{LAX}).

The corresponding elements $H_i,\ \ i=1,..,11$ are given by $H_i=h+a_i,$ where
\[\begin{array}{lclclclcl}
a_1&=&[u^{-1},vu]           &=&\ v\ \ (S^2I^{-1} - 1) ,       & \ & a_7&=&SI + a_2 + a_4 , \\
a_2&=&[v\ \ \ ,u^{-1}v^{-1}]&=&u^{-1}(\ \ \ I \ \ \ - 1) ,    & \ & a_8&=&SI + a_2 + [v,u^{-1}v^2] , \\
a_3&=&[v\ \ \ ,uv^{-1}]     &=&\ u\ \ (S^3I^{-1} - 1) ,       & \ & a_9&=&SI + a_4 + [u^{-1},u^{-1}v^{-1}u] ,\\
a_4&=&[u^{-1},v^{-1}u]      &=&v^{-1}(S\ \,\,I  \ \ \, - 1) , & \ & a_{10}&=&S^2I + [v^{-1},uv] , \\  
a_5&=&a_1+a_4 ,             & &                               & \ & a_{11}&=&\ \ \ \, I + [u,vu^{-1}]  \\
a_6&=&a_2+a_3 ,             & &                               & \ &       & & 
\end{array}\]
Here $S$ stands for a 
cyclic shift of factors in a monomial, i.e.\ $S(abc...z)=bc...za,$ and $I$ is given by (\ref{II}).  
Sets of $H_i, \overline{H_i}$ that are conjugate to each other are
$\{h, \overline{H_5}, \overline{H_6} \},$ 
$\{H_1, H_3, \overline{H_2}, \overline{H_4} \},$ 
$\{H_7\},$  $\{H_8,H_9,H_{10},H_{11}\}$ 
and the involuted versions of these 4 groups.   
In addition to $h, A$ three of the pairs $H_i, A_i$ representing these groups 
up to involution and conjugation are:
\[\begin{array}{ll}
H_1=u + u^{-1} +  v^{-1} +  u^{-1}v^{-1}+u^{-1} v u, & 
A_1= u - v + v^{-1} + u^{-1} v^{-1}, \\%T
H_7= u + v + u^{-1} v^{-1} u + v u^{-1} v^{-1} + u^{-1} v^{-1} + v u^{-1} v^{-1} u, &
A_7= v^{-1}-v, \\ %Y
%H_{9}= u + v + u^{-1} + u^{-1} v^{-1} u + v u^{-1} v^{-1} u + u^{-2} v^{-1} u, & 
%A_{9}= v + u^{-1} - v^{-1}-u^{-1} v^{-1}.%Z
H_{11}= u + u^{-1} + v^{-1} + u^{-1} v^{-1} + uvu^{-1} + uvu^{-1} v^{-1}, &
A_{11}= -u^{-1}. 
\end{array}\]

\section{The pre-Hamiltonian operator}

In the previous section we found several infinite sequences of trace
integrals for (\ref{inhom}).  Here we construct the sequences of
corresponding symmetries via a so called pre-Hamiltonian operator.
This operator is defined in terms of left and right multiplication
operators.

For any Laurent polynomial $a\in{\mathcal M}$ we denote by $L_a$ and
$R_a$ the operators of left and right multiplications on ${\mathcal M}$:
\[
  L_a (x)=a\,  x, \qquad  R_a (x)=x\,  a\, .
\]
It is clear that $L_{a  b} =L_a L_b$ and $R_{a  b} =R_b R_a.$
It follows from  the associativity of the algebra
${\cal M}$ that $R_a L_b=L_b R_a.$  

The algebra of all left and right multiplication operators is
generated by $L_{x_i},L_{x_{i}^{-1}},R_{x_i},R_{x_{i}^{-1}}$. We
denote this associative algebra by ${\cal O}$ and call it the 
{\it algebra of local operators}.

For any element $a=a({\bf x})\in {\cal M}$ we define an 
$1\times N$-matrix ${\bf a}_{*}$   with entries
being elements of ${\mathcal O}$ by the following identity:
\begin{equation} \label{frechdef}
 \frac{d}{d\epsilon}a({\bf x}+\epsilon\, \delta\! {\bf x})|_{\epsilon =0}=
{\bf a_{*}}(\delta \! {\bf x})\, .
\end{equation}
For example, for $h$ from (\ref{hh}) we have
\[ {\bf h_*}=( \ 1-L_{u^{-1}}R_{u^{-1}}-L_{u^{-1}}R_{u^{-1}}R_{v^{-1}} \ , \ \  
           1-L_{v^{-1}}R_{v^{-1}}-L_{u^{-1}}L_{v^{-1}}R_{v^{-1}} \ ) . \]
It is easy to see that 
\begin{equation} \label{frechet}
D_t(a)={\bf a_{*}}({\bf F}),
\end{equation}
where $D_t$ is the
derivation associated with (\ref{geneq}) and ${\bf F}=(F_1,...,F_N)^{T}$ 
is the right hand side of (\ref{geneq}).

For any vector ${\bf a}=(a_1,...,a_N)^{T},\,\,a_i\in {\cal M}$ we define the
Fr\'echet derivative operator ${\bf a}_{*}$ as the $N\times N$-matrix with
rows ${\bf (a_1)_*},...,{\bf (a_N)_*}$.

For any two vectors ${\bf p}=(p_1,...,p_N)^T,$    
${\bf q}=(q_1,...,q_N)^T,$  $p_i,q_i\in {\cal M}$ we put
\[ \langle {\bf p}, {\bf q}\rangle = p_{1}q_{1}+\cdots+p_{N} q_{N}. \] 
Let $a({\bf x})\in {\cal M}$. Then
$  \mbox{\em \bf grad}\, (a) $ is the vector uniquely defined by:
\[ \frac{d}{d\epsilon}a({\bf x}+\epsilon \, \delta \! {\bf x})|_{\epsilon
=0}\sim
\,\langle \delta \! {\bf x}, \mbox{\em \bf grad} \,( a({\bf x}))\rangle \, .\]
We will denote
by $\mbox{\em grad}_{x_{1}}(a),..., \mbox{\em grad}_{x_{N}}( a) $
the components of the vector \mbox{\em \bf grad}\,(a).
It is easy to see that if $a \sim b$, then $ \mbox{\em \bf grad}\, (a)= \mbox{\em \bf grad}\, (b)$. This
means that $ \mbox{\em \bf grad}\, (a) $ is well-defined for trace integrals.

For example, for the function $h$ given by (\ref{hh}) we have 
\[\mbox{\em \bf grad}\, h=( 1 - u^{-2} - u^{-1} v^{-1} u^{-1},\quad                   
                   1 - v^{-2} - v^{-1} u^{-1} v^{-1} )^T.\]

It follows from the definition of an infinitesimal symmetry (\ref{gensym}) 
with symmetry generator ${\bf G}=(G_{1},.., G_{N})^T$ 
for a system (\ref{geneq}) % with right hand site ${\bf F}$
that $D_t D_\tau {\bf x} = D_\tau D_t {\bf x}$, i.e.\
\begin{equation}
D_t {\bf G} = D_\tau {\bf F} = {\bf F}_*({\bf G}) \label{sycon}
\end{equation}
(from (\ref{frechet})) is a linearized equation satisfied by ${\bf G}$, 
where ${\bf F}_{*}$ is the Fr\'echet derivative of the right hand side of
(\ref{geneq}) (cf. \cite{olver}). 

An $N\times N$ matrix ${\mathcal P}$ with entries from ${\cal O}$ is
called a {\it pre-Hamiltonian operator} for equation (\ref{geneq}) if
\begin{equation}\label{hamop}
D_t({\mathcal P})={\bf F}_* {\mathcal P}+{\mathcal P} {\bf F}_* ^{\star}.
\end{equation}
Here  the adjoint operation $\star $ on ${\rm Mat}_N({\cal O})$ is uniquely defined by the formula 
\begin{equation}  \label{adj}
\langle {\bf p}, {\bf Q}({\bf q})\rangle \ \sim \ \langle {\bf Q}^{\star}({\bf p}), {\bf q} \rangle ,
\end{equation}
where  ${\bf Q} \in {\rm Mat}_N({\cal O}),$ $p_i,q_i\in {\cal M}.$
 
Relation (\ref{hamop}) can be rewritten in the form 
\begin{equation}\label{hamop1}
\left(D_t-{\bf F}_*\right){\mathcal P}={\mathcal P} \left(D_t+ {\bf F}_* ^{\star}\right).
\end{equation}
It can be shown (cf.\ \cite{olver}) that for any trace integral  
$a$ of (\ref{geneq}) the vector ${\bf b}=\mbox{\em \bf grad} (a)$  satisfies the equation 
$D_t({\bf b})+{\bf F}_* ^{\star}({\bf b})=0$. 
Applying both sides of (\ref{hamop1}) to ${\bf b}$, we get that any 
pre-Hamiltonian operator maps gradients of integrals for (\ref{geneq}) to
symmetries.

{\bf Proposition.} The following operator (cf. \cite{MikSok})
\begin{equation} \label{preHam}
{\mathcal P}=\left( \begin{array} {ccc} 
R_u R_u - L_u L_u                    \ \  & L_u L_v + L_u R_v - L_v R_u + R_u R_v \\
L_u R_v - L_v L_u - L_v R_u - R_v R_u \ \  & L_v L_v - R_v R_v
\end{array} \right)
\end{equation}
is a pre-Hamiltonian operator for equation (\ref{inhom}). 
The proof consists of the straightforward verification of relation (\ref{hamop}) using 
\[{\bf F}_* =  \left( \begin{array} {ccc} 
R_v-R_{v^{-1}} \ \ & L_u + L_u L_{v^{-1}} R_{v^{-1}} + L_{v^{-1}} R_{v^{-1}} \\
- L_v - L_v L_{u^{-1}} R_{u^{-1}} - L_{u^{-1}} R_{u^{-1}} \ \ & - R_u + R_{u^{-1}} 
\end{array} \right) \]
\[{\bf F}_*^{\star} = \left( \begin{array} {ccc} 
L_v-L_{v^{-1}} \ \ & - R_v - L_{u^{-1}} R_{u^{-1}} R_v - L_{u^{-1}} R_{u^{-1}} \\
R_u + L_{v^{-1}} R_{v^{-1}} R_u + L_{v^{-1}} R_{v^{-1}} \ \ & - L_u + L_{u^{-1}} 
\end{array} \right) \]
computed from (\ref{frechdef}), (\ref{adj}).

Applying operator (\ref{preHam}) to $\mbox{\em \bf grad}\, h, $ we get (up to
a factor of 2) the right hand side of system (\ref{inhom}). In this sense
$h/2$ plays the role of a Hamiltonian for (\ref{inhom}). However, the
bracket
\[ \{a,b\} = <\mbox{\em \bf grad}\,a , \, {\mathcal P}\, \mbox{\em \bf grad}\,b > \]
defined on traces of Laurent polynomials does not satisfy the Jacobi
identity for ${\mathcal P}$ from (\ref{preHam}) 
and a true Hamiltonian structure for (\ref{inhom}) is yet unknown.
 
\section{The Lax pair} \label{LaxPair}

 Consider 
\begin{equation}\label{LA} {\bf L} = \left(
\begin{array}{cc}
  v^{-1} + u \ &
    \lambda v + v^{-1} u^{-1} + u^{-1} + 1 \\
   v^{-1} + \frac{1}{\lambda} u \ &
   v + v^{-1}u^{-1} + u^{-1} + \frac{1}{\lambda}
\end{array} \right) , \qquad 
{\bf A} = \left(
\begin{array}{cc}
   v^{-1} - v + u \    &
   \lambda v  \\
   v^{-1}         \    &
   u
\end{array}   \right),
\end{equation}
where $\lambda$ is a (scalar) spectral parameter.
Then the relation  
\[D_t {\bf L} = [{\bf A},{\bf L}]\]
is equivalent to the Kontsevich system (\ref{inhom}). Although the pairs
$H_i,A_i$ described in Section \ref{SymInt} do also satisfy an equation
similar to (\ref{LAX}), they are not Lax pairs for (\ref{inhom}) since in this
case (\ref{LAX}) follows from (\ref{inhom}) but not vice versa.

The Lax pair can be replaced by any equivalent one obtained through a
conjugation (\ref{gauge}), where $g$ is an arbitrary invertible Laurent $2\times
2$-matrix. Other equivalence transformations are ${\bf L}\to P_1({\bf L}),\,\, {\bf A}\to
{\bf A}+P_2({\bf L}),$ where $P_i$ are polynomials with constant $\lambda$-dependent
coefficients, and arbitrary transformation $\lambda\to f(\lambda).$

As usual, the traces $tr \ {\bf L}^m = ({\bf L}^m)_{11}+({\bf L}^m)_{22}$
generate trace integrals of motion. In particular, 
$tr \ {\bf L}$ yields $v^{-1} + u+ v^{-1} u^{-1} + u^{-1}+ v$, which is equivalent to $h$
from (\ref{hh}). In contrast to (\ref{LAX}) each power of ${\bf L}$ gives us several
trace integrals since $tr \ {\bf L}^m$ is a polynomial in $\lambda, \lambda^{-1}$
with all coefficients being trace integrals. We verified that all trace integrals of
degree $\le 12$ for (\ref{inhom}) are generated in such a way.

Table \ref{table1b} shows all integrals of degree $d$ generated from $tr \ {\bf L}^m, \ \ m\leq 14$
that are not generated from $tr \ {\bf L}^i, \ i<m$. The following statements assume that 
all trace integrals have been reduced modulo lower degree trace integrals.

Each integral is represented by a $\star$, $\circ$ or $\bullet$ and is located
in one diagonal of table \ref{table1b}. Each diagonal starts in a table entry
which shows a number $k$ indicating that the integral is $I^k$. For the
single $\star$ - diagonal is $k=0$.  $\circ$ - diagonals have $k<0$ and start in
row $3|k|$, i.e.\ $I^k$ result from $tr \ {\bf L}^{3|k|}$ and $\bullet$ - diagonals 
have $k>0$ and start in row $4k$, i.e.\ $I^k$ result from $tr \ {\bf L}^{4k}$.

The different powers of $\lambda$ in $tr \ {\bf L}^m$ have the following first 
integrals as coefficients.  $\star$ - integrals are the coefficients of
$\lambda^0$.  $\circ$ - integrals of degree $d$ resulting from ${\bf L}^m$ are
the coefficients of $\lambda^{(d-2m)/2}$.  $\bullet$ - integrals of
degree $d$ resulting from ${\bf L}^m$ are the coefficients of $\lambda^{(2m-d)/4}$.

All $\star$ - integrals are invariant under involution (\ref{invol}).
All other first integrals come in pairs, one $\circ$ - and one
$\bullet$ - integral, both are involution symmetric to each other and
therefore in the same column. 

Applying the involution (\ref{invol}) to the Lax pair (\ref{LA}), we get a dual one: 
\begin{equation} \label{invLA} \overline{{\bf L}} = \left(
\begin{array}{cc}
  u^{-1} + v \ &
   \lambda u + u^{-1} v^{-1} + v^{-1} + 1 \\
   u^{-1} + \frac{1}{\lambda} v \ &
   u + u^{-1} v^{-1} + v^{-1} + \frac{1}{\lambda}
\end{array} \right) , \qquad 
\overline{{\bf A}} = \left(
\begin{array}{cc}
   u^{-1} - u + v \    &
   \lambda u  \\
   u^{-1}         \    &
   v
\end{array}   \right). \end{equation}
%It is possible to verify that the dual Lax pair is not equivalent to (\ref{LA})
%but each one of ${\bf L}$ and $\overline{{\bf L}}$ generates the same vector 
%space of first integrals, verified up to degree 14.

We tried to find a $2\times 2$ Laurent matrix ${\bf P}$ such that
${\bf P\, \overline{L} = {\rm const}\, L\, P}$ (under change of $\lambda$)
but we did not succeed and we doubt such a ${\bf P}$ exists.
Therefore we think that ${\bf L}$ and ${\bf \overline{L}}$ 
are not related by standard algebraic 
symmetries. ${\bf L, A}$ and ${\bf \overline{L}, \overline{A}}$
are different Lax pair representations
 but each one gives all trace first integrals, only for
 different degrees. For example, see table 1,
$tr \ {\bf L}^3$ generates $I^{-1}$ as first integral and
$tr \ {\bf \overline{L}}^3$ generates $I$ as first integral.
${\bf L}$ does also generate $I$ as first integral, but from $tr \ {\bf L}^4$
not from $tr \ {\bf L}^3$. And that is the case with each first integral.
Using $\overline{{\bf L}}$ instead of ${\bf L}$ does
generate the same table only with $\circ$ and $\bullet$ - entries interchanged.
Replacing $\lambda \to f(\lambda)$ in ${\bf L}$ (\ref{LA}) does not change the
table. 

Applying the pre-Hamiltonian operator (\ref{preHam}) to $\star$ -
integrals gives involution invariant symmetries of the same degree.
Applying (\ref{preHam}) to an involution symmetric pair of first
integrals gives 2 symmetries that are one degree higher than the first
integral and that are also involution symmetric to each other.

Traces of $H_j^k$ (\ref{Hjk}) have been verified to be linear combinations 
of integrals of table \ref{table1b}. For example,
\begin{itemize}
\item first integrals $tr \ h^{2m+k}, \ n=0,1,..,\ k=0,1$
are of degree $2(2m+k)$ and require only up to ${\bf L}^{2m+k}$ to be derived,
\item first integrals $tr(h+a_1)^{2m+k}, \ m=0,1,..,\ k=0,1$
are also of degree $2(2m+k)$ (although $h+a_1$ is of degree 3)
but require up to ${\bf L}^{4m+k}$ to be derived,
\item first integrals $tr(h+a_7)^m$ and $tr(h+a_{11})^m, \ m=0,1,..$
are of degree $4m$ and require up to ${\bf L}^{3m}$ to be derived.
\end{itemize}

All first integrals generated by the Lax pair operator $\overline{{\bf L}}
(= {\bf L}|_{u\leftrightarrow v})$ are linear combinations of first
integrals generated from ${\bf L}$ and vice versa.

If a given first integral $F$ of degree $2(2m+k), \ m=0,1,...,\ k=0,1$ is to
be expressed as a linear combination of first integrals computed from ${\bf L}$ and
$\overline{{\bf L}}$ then it requires at least ${\bf L}^{2m+k}$ or $\overline{{\bf L}}^{2m+k}$
and at most ${\bf L}^{3m+k}$ and $\overline{{\bf L}}^{3m+k}$. If $F$ is to be expressed
as a linear combination of first integrals computed from ${\bf L}$ alone then it
requires at most ${\bf L}^{4m+k}$.

The table of first integrals with its straightforward extension
appears to be complete because all symmetries of degree up to 16 have
been verified to be generated from integrals of this table and  
the pre-Hamiltonian operator (\ref{preHam}).

\begin{table}
   \begin{center}
   \begin{tabular}{|c|c|c|c|c|c|c|c|c|c|c|c|c|c|c|c|} \hline
    $m \backslash d$&\ 0\ &\ 2\ &\ 4\ &\ 6\ &\ 8\ &10&12&14&16&18&20&22&24&26&28 \\ \hline
      0  &$\star^0$& & & & &  &  &  &  &  &  &  &  &  &   \\ \hline
      1  & &$\star$& & & & &  &  &  &  &  &  &  &  &   \\ \hline
      2  & & &$\star$& & & &  &  &  &  &  &  &  &  &   \\ \hline
      3  & & &$\!\!\circ^{-1}\!\!\!$&$\star$& &  &  &  &  &  &  &  &  &  &   \\ \hline
      4  & & &$\!\!\bullet^{+1}\!\!\!$&$\circ$&$\star$&  &  &  &  &  &  &  &  &  &   \\ \hline
      5  & & & &$\bullet$&$\circ$&$\star$&  &  &  &  &  &  &  &  &   \\ \hline
      6  & & & & &$\!\!\bullet\circ^{\!\!-2}\!\!\!$&$\circ$&$\star$&  &  &  &  &  &  &  &   \\ \hline
      7  & & & & & &$\bullet\circ$&$\circ$&$\star$&  &  &  &  &  &  &   \\ \hline
      8  & & & & &$\!\!\bullet^{+2}\!\!$&  &$\bullet\circ$&$\circ$&$\star$&  &  &  &  &  &   \\ \hline
      9  & & & & & &$\bullet$&$\circ^{-3}\!\!\!$&$\bullet\circ$&$\circ$&$\star$&  &  &  &  &   \\ \hline
     10  & & & & & &  &$\bullet$&$\circ$&$\bullet\circ$&$\circ$&$\star$&  &  &  &   \\ \hline
     11  & & & & & &  &  &$\bullet$&$\circ$&$\bullet\circ$&$\circ$&$\star$&  &  &   \\ \hline
     12  & & & & & &  &$\!\bullet^{+3}\!\!\!$&  &$\!\!\bullet\circ^{\!-4}\!\!\!$&$\circ$&$\bullet\circ$&$\circ$&$\star$&  &   \\ \hline
     13  & & & & & &  &  &$\bullet$&  &$\bullet\circ$&$\circ$&$\bullet\circ$&$\circ$&$\star$&   \\ \hline
     14  & & & & & &  &  &  &$\bullet$&  &$\bullet\circ$&$\circ$&$\bullet\circ$&$\circ$&$\star$ \\ \hline
   \end{tabular}
   \end{center}
   \caption{{\rm \# of new trace first integrals of degree $d$ generated from
       $tr \ {\bf L}^m$ not generated from $tr \ {\bf L}^i, \ i<m$.}}
  \label{table1b}
\end{table}  
%--------------------------
\section{Summary}

In this paper we provide two Lax pair representations ${\bf L},
{\bf A}$ in (\ref{LA}) and $\overline{{\bf L}},\overline{{\bf A}}$ in
(\ref{invLA}) for the non-abelian Kontsevich ODE system
(\ref{inhom}). We also give a pre-Hamiltonian operator ${\mathcal P}$
in (\ref{preHam}) that converts gradients of first integrals into Lie
symmetries. By computing independently all symmetries with Laurent
polynomial generators up to degree 16 we show that both Lax pairs
generate all trace first integrals at least up to this degree.
For any $n$ we propose a conjecture on the dimension of the vector 
space of symmetries of order $\le n$.

Outstanding work contains an algebraic description of all 
${\bf A}$-operators for the hierarchy and a recursion operator 
for the model.

Another open problem is to find a Laurent Lax
pair for the original discrete Kontsevich map.
We have checked numerically  for $2\times 2$ and $3\times 3$ matrices
$u,v$ that the characteristic polynomial of
${\bf L}(\lambda)$ does not change under the discrete map
(\ref{descr}). In this case we can apply the
discrete map in components, it is invertible. 
% 1. from me
%  But that is not the case on the level of Laurent polynomials since we
%  cannot apply the map to $L$.  It is an outstanding problem to find a
%  conjugation Laurent matrix ${\bf P}$ such that the discrete map is
%  applicable to ${\bf P L P^{-1}}$.
% 2. from  me
%  But that is not the case on the level of Laurent polynomials since
%  applying the map to the components of ${\bf L}$ does not produce
%  Laurent polynomials. What is needed is a conjugation map ${\bf P}$
%  such that the discrete map applied to ${\bf P L P^{-1}}$ gives a
%  matrix of Laurent polynomials that is also conjugate to ${\bf L}$.
% 3. next from Volodya
But this is not the case on the level of Laurent polynomials since we
cannot apply the map to $L$ given in (\ref{LA}). It is an outstanding
problem to find a Laurent matrix ${\bf P}$ such that the discrete map is
applicable to $\tilde L = P L P^{-1}$.

\vskip.3cm \noindent {\bf Acknowledgements.} 
Authors thank M.\ Kontsevich and V.\ Sokolov for fruitful discussions.
E.\ Schr\"{u}fer is thanked for discussions and computer algebra support.
O.E. is grateful to IHES for hospitality. She was partially supported by 
the RFBI grant 11-01-00341-a. Computations were run on computer hardware 
of the Sharcnet consortium (www.sharcnet.ca).

%\newpage

\end{document}